# Microstructural Control and Heat Transport Enhancement in Lanthanum Sulfate for Thermochemical Heat Storage

Author names

Kunihiko Shizume[1,2*], Naoyuki Hatada[1*]

* Corresponding Author

Affiliations

[1] Kyoto University

[2] Nagoya Institute of Technology

Address:

Yoshida-honmachi, Sakyo-ku, Kyoto city, Kyoto 606-8501, Japan

Email address:

shizume@nitech.ac.jp

hatada.naoyuki.8u@kyoto-u.ac.jp

K. Shizume's current affiliation: Nagoya Institute of Technology.

Current address: Gokiso-cho, Showa-ku, Nagoya, Aichi, 466-8555, Japan

**Abstract**

Enhancing heat transport within thermochemical heat storage (TCHS) materials is essential for improving their heat output. A common strategy is to combine salts with highly thermally conductive additives, such as carbon or metallic materials. However, such composites often exhibit interfacial instability and reduced gas permeability. In this work, we propose an alternative approach based on microstructural orientation control, aiming to create efficient heat-transport pathways without relying on conductive additives. β-$La_2(SO_4)_3$, which undergoes reversible hydration and dehydration below 250 °C, was selected as a model TCHS material. Highly oriented rod-like $La_2(SO_4)_3\cdot 9H_2O$ crystals with centimeter-scale lengths were grown from solution, cut into plate-shaped specimens, and then dehydrated to β-$La_2(SO_4)_3$. Two types of specimens with different microstructural orientations, which form spontaneously during the dehydration of $La_2(SO_4)_3\cdot 9H_2O$ to β-$La_2(SO_4)_3$, were prepared. In the "cross-plane-gb" specimen, the aligned grain boundaries were predominantly oriented parallel to the through-thickness direction of the plate, whereas in the "in-plane-gb" specimen, they were predominantly oriented perpendicular to this direction. Laser flash analysis (LFA) of β-$La_2(SO_4)_3$ revealed a clear orientation dependence of heat transport: the apparent thermal diffusivity was approximately 0.24 $mm^2/s$ for the cross-plane-gb specimens, in which the grain boundaries are aligned along the heat-flow direction during LFA, and it was approximately 0.15 $mm^2/s$ for the in-plane-gb specimens. These findings demonstrate that controlling the microstructural orientation is a viable route for enhancing heat transport in TCHS materials, offering an additive-free design strategy.

**Keywords**: Thermochemical heat storage; Lanthanum sulfate; Microstructure orientation; Grain boundary; Thermal diffusivity; Laser flash analysis

## 1. Introduction

Waste heat recovery using thermal energy storage is crucial for improving energy efficiency, given that more than half of the primary energy supply is lost as waste heat in the world's energy-conversion chain [1–3]. In particular, thermochemical heat storage (TCHS) is a promising technology owing to its long-term storage ability and high energy density compared with sensible and latent heat storage [4,5]. However, no widespread commercial TCHS applications have been reported to date, although numerous potential reaction systems involving gas–gas, gas–liquid, and gas–solid reactions have been studied [6,7]. Gas–solid reactions using inorganic compounds as solid reactants are particularly attractive [8–10], such as $MgO/H_2O$ [11,12], $CaO/H_2O$ [13], $CaCl_2/NH_3$ [14], and $CaO/CO_2$ [15,16]. Inorganic reactants are typically shaped into gas-permeable bodies, such as those with a porous structure, to ensure sufficient gas transport and enhance reactivity in practical applications [17,18]. However, when these materials are formed into porous compacts or pellets, their inherently poor interparticle connectivity results in a significantly reduced effective thermal conductivity (TC), generally ranging from 0.1 to 0.4 W $m^{-1}$ $K^{-1}$ [19,20]. By contrast, dense bodies of inorganic compounds exhibit much higher TC values, including MgO (40–60 W $m^{-1}$ $K^{-1}$) and CaO (15 W $m^{-1}$ $K^{-1}$) [21,22]. Such low thermal conductivity in porous structures can hinder heat transport and limit the overall reaction kinetics within the reactant body [23,24].

Therefore, enhancing heat transport within the reactant body is essential for improving the performance of gas–solid TCHS systems. A common strategy is to combine inorganic compounds with highly thermally conductive additives, such as carbon or metallic materials [25–27]. However, without a careful optimization of material selection and combining methods, this strategy often faces challenges such as low cyclic stability owing to unstable compound–additive interfaces and suppressed gas permeability resulting from reduced porosity [28,29]. In this work, as an alternative approach, we did not rely on composite formation but instead attempted to enhance the heat transport of the compound body itself through microstructural control.

We explore the β phase of $La_2(SO_4)_3$, which formally corresponds to $\beta\text{-}La_2(SO_4)_3{\cdot}xH_2O$ ($0 \leq x \leq 1$), as a model TCHS material that undergoes the following reversible reaction (Equation (1)) in the temperature range of approximately 80–220 °C:

$$\beta\text{-}La_2(SO_4)_3\ (s) + H_2O\ (g) \rightleftarrows \beta\text{-}La_2(SO_4)_3{\cdot}xH_2O\ (s), \quad 0 \leq x \leq 1 \qquad (1)$$

This material has a moderate heat storage capacity; the standard enthalpy change associated with this reaction between hydration numbers $x = 1$ and $x = 0$ is 91 ± 13 kJ $mol^{-1}$ [30]. Although the hydration number, $x$, varies depending on the ambient conditions, we hereafter refer to this phase simply as $\beta\text{-}La_2(SO_4)_3$. $\beta\text{-}La_2(SO_4)_3$ has been proposed as a candidate for low-temperature TCHS (<250 °C) owing to its favorable thermodynamics and a rapid hydration reaction rate [30–32].

One origin of this rapid reaction rate is the characteristic microstructure that facilitates $H_2O$ diffusion [31]. This microstructure of $\beta\text{-}La_2(SO_4)_3$ is known to form spontaneously upon the dehydration of the $La_2(SO_4)_3{\cdot}9H_2O$ precursor.

When $La_2(SO_4)_3 \cdot 9H_2O$ is synthesized from aqueous solution, it crystallizes in the hexagonal crystal system and forms rod-like crystals whose longitudinal direction is parallel to the c-axis, without any pronounced fine microstructure. We experimentally confirmed that the grain boundaries in this microstructure of β-$La_2(SO_4)_3$ were predominantly aligned along the longitudinal direction of the rod-like crystals. The grain boundaries have been confirmed to function as efficient diffusion pathways for $H_2O$[31]. In addition, the hydration/dehydration reaction of β-$La_2(SO_4)_3$ shows no noticeable degradation in reaction rate or capacity even after 100 cycles[30]. This stability suggests that both the microstructure and the phase of β-$La_2(SO_4)_3$ are well maintained during repeated cycling. In this study, we further explore the possibility that this microstructure also functions as an effective heat transport pathway. Because grain boundaries are preferentially aligned along the longitudinal direction, heat transport parallel to this direction involves fewer boundary crossings and is therefore expected to reduce grain boundary thermal resistance.

In this study, highly oriented rod-like $La_2(SO_4)_3 \cdot 9H_2O$ crystals with centimeter-scale lengths were grown from a saturated aqueous solution, where the longitudinal direction is parallel to the crystallographic c-axis. The crystals were cut into plate-shaped specimens with different orientations relative to the longitudinal direction and then dehydrated to β-$La_2(SO_4)_3$. This procedure produced two types of plates: "cross-plane-gb" and "in-plane-gb" specimens, in which the grain boundaries were predominantly aligned along the cross-plane direction, corresponding to the through-thickness direction, and along the in-plane direction, respectively. The thermal diffusivity in the cross-plane direction was measured using laser flash analysis (LFA). The cross-plane-gb specimen showed a higher apparent thermal diffusivity (~0.24 $mm^2/s$) than the in-plane-gb specimen (~0.15 $mm^2/s$ ). These results highlight microstructural orientation control as a practical route for improving heat transfer in TCHS materials.

## 2. Methodology

### 2.1 Solubility study of $La_2(SO_4)_3$ by thermogravimetry

The solubility of $La_2(SO_4)_3$ in water was measured to provide essential data for controlling the growth of $La_2(SO_4)_3 \cdot 9H_2O$ crystals. Although the solubility was previously reported [33,34], we re-examined it under our experimental conditions.

To prepare a saturated aqueous solution of $La_2(SO_4)_3$ at room temperature, 20 g of $La_2(SO_4)_3 \cdot 9H_2O$ (>98.0%, Wako Pure Chemical, Osaka, Japan) was first dissolved in 2 L of deionized water at 60 °C, then cooled and maintained at room temperature. Under these conditions, gradual evaporation of water induced the precipitation of $La_2(SO_4)_3 \cdot 9H_2O$, and the remaining solution became saturated with respect to $La_2(SO_4)_3$. This room-temperature saturated solution was used as the initial solution to investigate the temperature dependence of solubility, which decreases with increasing temperature, as reported by Muthmann [33]. The saturated solution was then distributed into sealed vessels and maintained at 30, 35, 40, 55, and 70 °C until precipitation from the supersaturated state was complete. To facilitate this process, a small amount of $La_2(SO_4)_3 \cdot 9H_2O$ powder was also added as seed crystals.

Solution samples were collected after varying holding times (10, 48, 100, 201, and 432 h). The amounts of solvent and solute were quantified by thermogravimetric analysis (TGA; Rigaku TG8120). Each solution sample was heated under dry argon (Ar) flow at 80 °C until it dried completely and converted to solid $La_2(SO_4)_3 \cdot 9H_2O$. From the initial mass and the mass of the dried residue, the mass fractions of water and $La_2(SO_4)_3 \cdot 9H_2O$ in the solution were determined. The temporal change of residual $La_2(SO_4)_3$ mass was used to determine the equilibrium solubility. The TGA results are presented in Fig. S1 (Supplementary Material).

### 2.2 Sample preparation for laser flash analysis

#### 2.2.1 Crystal growth of $La_2(SO_4)_3 \cdot 9H_2O$

Highly oriented rod-like $La_2(SO_4)_3 \cdot 9H_2O$ crystals were grown from aqueous solution using the setup illustrated in Fig. 1. The apparatus consisted of a reservoir containing saturated solution at ~21 °C (room temperature) and a crystal-growth bath maintained at 40–70 °C using a mantle heater. As shown in Section 3.1, the solubility decreases with increasing temperature. Circulation of the solution between the two vessels using a liquid-transfer pump ensured supersaturation in the hot bath and promoted crystal growth. A cartridge filter was installed in the line to prevent accidental transfer of solids. This process was continued for several months, until centimeter-sized crystals were obtained.

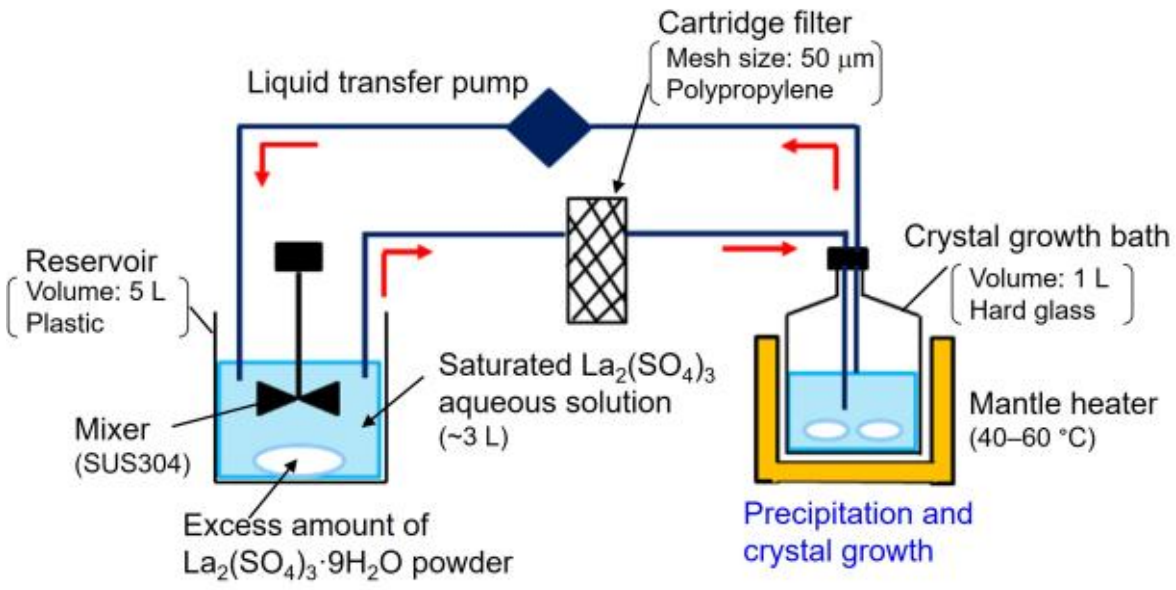

**Fig. 1.** Schematic of the experimental setup for crystal growth of $La_2(SO_4)_3 \cdot 9H_2O$. The reservoir was a 5 L plastic container filled with ~3 L of saturated $La_2(SO_4)_3 \cdot 9H_2O$ solution, into which excess $La_2(SO_4)_3 \cdot 9H_2O$ solid was also introduced to maintain supersaturation. The crystal-growth bath was a 1 L glass vessel heated and maintained at 40–70 °C using a mantle heater. The two vessels were connected by tubing with a liquid-transfer pump to circulate the solution.

#### 2.2.2 Shaping of specimens

The highly oriented rod-like $La_2(SO_4)_3 \cdot 9H_2O$ crystals were processed into plate-shaped specimens with dimensions suitable for LFA measurements. Owing to the hexagonal crystal structure, the highly oriented $La_2(SO_4)_3 \cdot 9H_2O$ crystals adopted a rod-like morphology, with their longitudinal direction parallel to the c-axis. Following dehydration from $La_2(SO_4)_3 \cdot 9H_2O$ to β-$La_2(SO_4)_3$, a fine microstructure developed, in which the grain boundaries were aligned approximately along this longitudinal direction. Based on this relationship, each rod-like crystal was cut into plate-shaped specimens such that the through-thickness direction was either parallel or perpendicular to the longitudinal direction of the precursor crystal. After dehydration, the resulting plates corresponded to specimens in which the grain boundary (gb) alignment in β-$La_2(SO_4)_3$ was predominantly parallel or perpendicular to the through-thickness direction. Hereinafter, the precursor $La_2(SO_4)_3 \cdot 9H_2O$ plates and the dehydrated β-$La_2(SO_4)_3$ plates are referred to as the “cross-plane-gb” and “in-plane-gb” specimens, respectively, according to the expected grain-boundary alignment after dehydration.

The specimens were then trimmed to dimensions of 6 × 6 mm (width × height) thickness of several millimeters using a diamond cutter and alumina abrasive paper. Depending on the experiment, the specimens were dehydrated to β-$La_2(SO_4)_3$ either *in situ* during heating in the LFA apparatus or *ex situ* in a furnace prior to LFA measurements. The crystallographic orientation of the precursor plates was determined using X-ray diffraction (XRD; X’Pert PRO MPD diffractometer with a Cu Kα radiation source, PANalytical, Almelo, Netherlands).

#### 2.2.3 Preparation of β-$La_2(SO_4)_3$ powder compact

A powder compact of β-$La_2(SO_4)_3$ was prepared as a randomly oriented specimen. $La_2(SO_4)_3 \cdot 9H_2O$ powder was ball-milled with yttria-stabilized zirconia balls (Tosoh, YTZ) in 2-propanol for 24 h and then heated at 300 °C for 90 min to obtain β-$La_2(SO_4)_3$. The resulting powder was uniaxially pressed into a 10-mm-diameter pellet with a relative density of 60%, which was estimated from the geometric density. A 6 × 6 mm plate-shaped specimen was cut from the pellet using a diamond cutter and alumina abrasive paper.

### 2.3 Measurements

Thermal diffusivity was measured using a laser flash apparatus (LFA467, NETZSCH, Selb, Germany). The cross-plane-gb and in-plane-gb specimens of $La_2(SO_4)_3 \cdot 9H_2O$ were coated with a Pt-sputtered layer to block light transmission and subsequently sprayed with carbon to ensure absorptivity during measurement. The specimens were subsequently heated to 300 °C, during which $La_2(SO_4)_3 \cdot 9H_2O$ dehydrated to β-$La_2(SO_4)_3$. To mitigate cracking during dehydration, the side surfaces of the cross-plane-gb and in-plane-gb specimens for *ex situ* tests were reinforced with inorganic putty (refractory putty HJ-112, Cemedine, Tokyo, Japan) or silicone rubber (RTV silicone rubber KE-3418, ShinEtsu Silicone, Tokyo, Japan) before heating. Pre-dehydrated cross-plane-gb and in-plane-gb specimens of β-$La_2(SO_4)_3$ were prepared in a furnace. The pre-dehydrated cross-plane-gb and in-plane-gb specimens and the powder compact of β-$La_2(SO_4)_3$ were measured by LFA. The cross-sectional microstructure of selected specimens after LFA was analyzed using scanning electron microscopy (SEM) with electron probe microanalysis (EPMA; JXA-8530F, JEOL, Tokyo, Japan).

## 3. Results and Discussion

### 3.1 Solubility study of $La_2(SO_4)_3$

The solubility of $La_2(SO_4)_3$ was systematically investigated to obtain fundamental data for determining the operating conditions of crystal growth. Fig. 2 shows the obtained solubility as a function of temperature. The values were determined by quantifying the dissolved mass of $La_2(SO_4)_3$ in saturated solutions at varying temperatures using TGA, as explained in Fig. S1. For comparison, the solubility data reported by Muthmann [33] are also plotted in Fig. 2. The reasonable agreement between the two sets of data confirms the reliability of our measurements. Because $La_2(SO_4)_3$ exhibits a decreasing solubility trend with increasing temperature, the crystal-growth bath was maintained at a higher temperature than the reservoir to ensure a continuous driving force for precipitation.

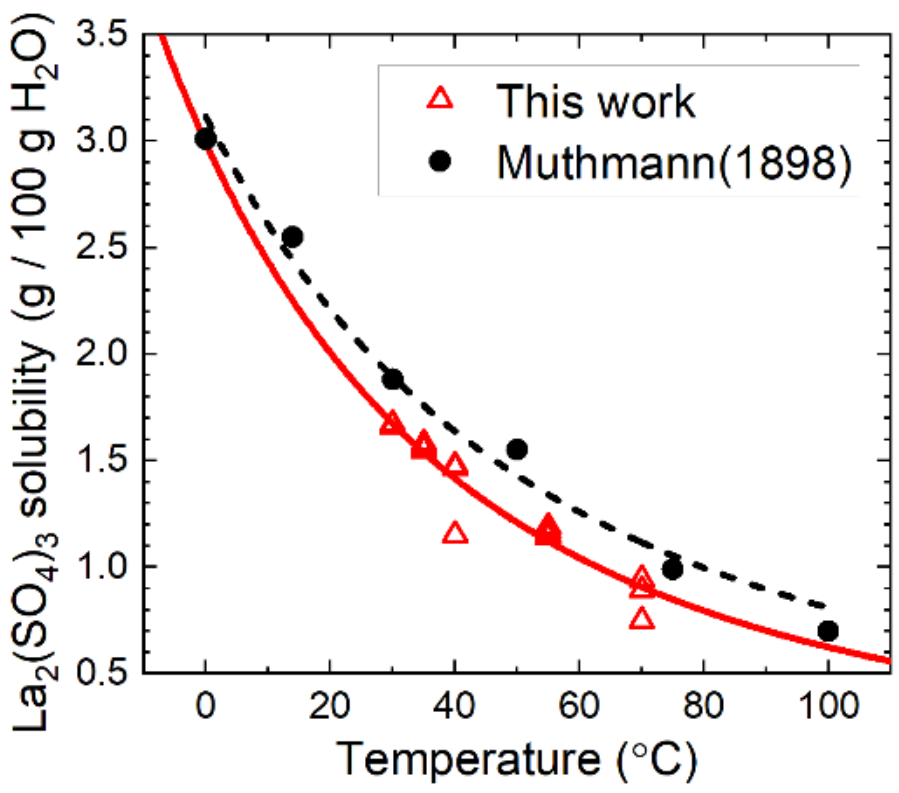

**Fig. 2.** Temperature dependence of the solubility of $La_2(SO_4)_3$. The red open triangles represent the solubilities at 30, 35, 40, 55, and 70 °C, as obtained in this study. These values were determined by preparing saturated solutions at each temperature and quantifying the dissolved solute by TGA (see Fig. S1). The black circles denote the literature data reported by Muthmann [33]. Both the present and literature data were well described by the van't Hoff equation, as shown by the red solid and black dashed lines, respectively.

### 3.2 Crystal growth of $La_2(SO_4)_3 \cdot 9H_2O$ and preparation of cross-plane-gb and in-plane-gb specimens

Through crystal-growth experiments conducted over several months, sufficiently large precursor crystals of $La_2(SO_4)_3 \cdot 9H_2O$ were obtained. Figs. 3(a) and 3(b) show the specimen preparation for cross-plane-gb and in-plane-gb specimens. Panels (a-i) and (b-i) schematically illustrate the geometric relationship between the plate-shaped specimens and the as-grown precursor crystals of $La_2(SO_4)_3 \cdot 9H_2O$. For the cross-plane-gb and in-plane-gb specimens, the through-thickness direction of the plate is parallel and perpendicular, respectively, to the longitudinal direction of rod-like $La_2(SO_4)_3 \cdot 9H_2O$ crystals. Panels (a-ii) and (b-ii) display the as-grown $La_2(SO_4)_3 \cdot 9H_2O$ crystals with centimeter-scale length. $La_2(SO_4)_3 \cdot 9H_2O$ crystallizes in the hexagonal crystal system from the aqueous solution, and the longitudinal direction of the rod-like crystals corresponds to the crystallographic c-axis, as confirmed by XRD analysis (Fig. 4). To prepare cross-plane-gb and in-plane-gb specimens, the crystals were sectioned using a diamond cutter in two different orientations and then mechanically polished to form plate-shaped specimens for LFA measurements, with lateral dimensions of approximately 6 × 6 mm and a thickness of several millimeters, as shown in panels (a-iii) and (b-iii).

The crystallographic orientations of the $La_2(SO_4)_3{\cdot}9H_2O$ plates for cross-plane-gb and in-plane-gb specimens were further examined using XRD. Fig. 4 shows the measurement setup and diffraction patterns. The specimens were mounted on a zero-background silicon plate and θ–2θ scans were performed on the plate surfaces, as illustrated in Fig. 4(a) and 4(b). The resulting diffraction patterns are presented in Fig. 4(c), together with the reference pattern of $La_2(SO_4)_3{\cdot}9H_2O$ from the International Centre for Diffraction Data. For the cross-plane-gb specimen, strong diffractions were observed from the (002) and (004) planes, indicating that the c-axis is predominantly parallel to the cross-plane direction. This also supports that the c-axis is aligned with the longitudinal direction of the rod-like $La_2(SO_4)_3{\cdot}9H_2O$ crystals. For the in-plane-gb specimen, strong diffractions were observed from the (100) and (140) planes, indicating that the c-axis lies approximately in-plane with respect to the plate surface. Although weak diffractions from non-preferred orientations were also present, the dominant diffractions verified that the prepared specimens were polycrystalline but preferentially oriented. Based on this specimen geometry and preferred orientation, the grain-boundary alignment that develops upon dehydration to β-$La_2(SO_4)_3$ is expected to be oriented either across the specimen thickness (cross-plane) or within the plate plane (in-plane). In the following sections, the cross-sectional microstructures of these two specimen types are examined, and their effects on heat transport are evaluated.

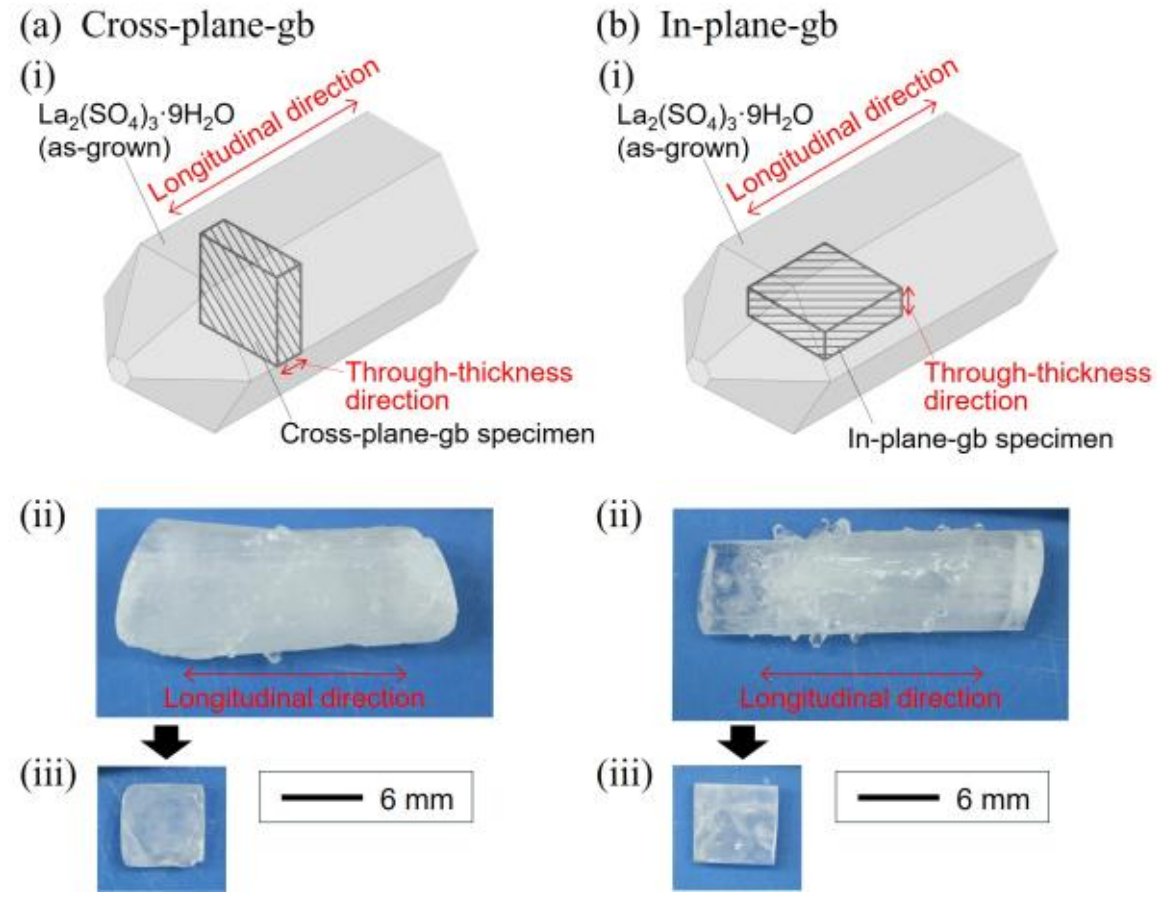

**Fig. 3.** Representative images of the grown crystals and the preparation processes of (a) cross-plane-gb and (b) in-plane-gb specimens. In both cases, (i) shows schematic of geometric relationships between plate-shaped specimens of β-$La_2(SO_4)_3$ and as-grown rod-like $La_2(SO_4)_3{\cdot}9H_2O$ crystal. (ii) shows the as-grown crystals, and (iii) shows the final plate-shaped specimens with lateral dimensions of approximately 6 × 6 mm after polishing with alumina abrasive paper.

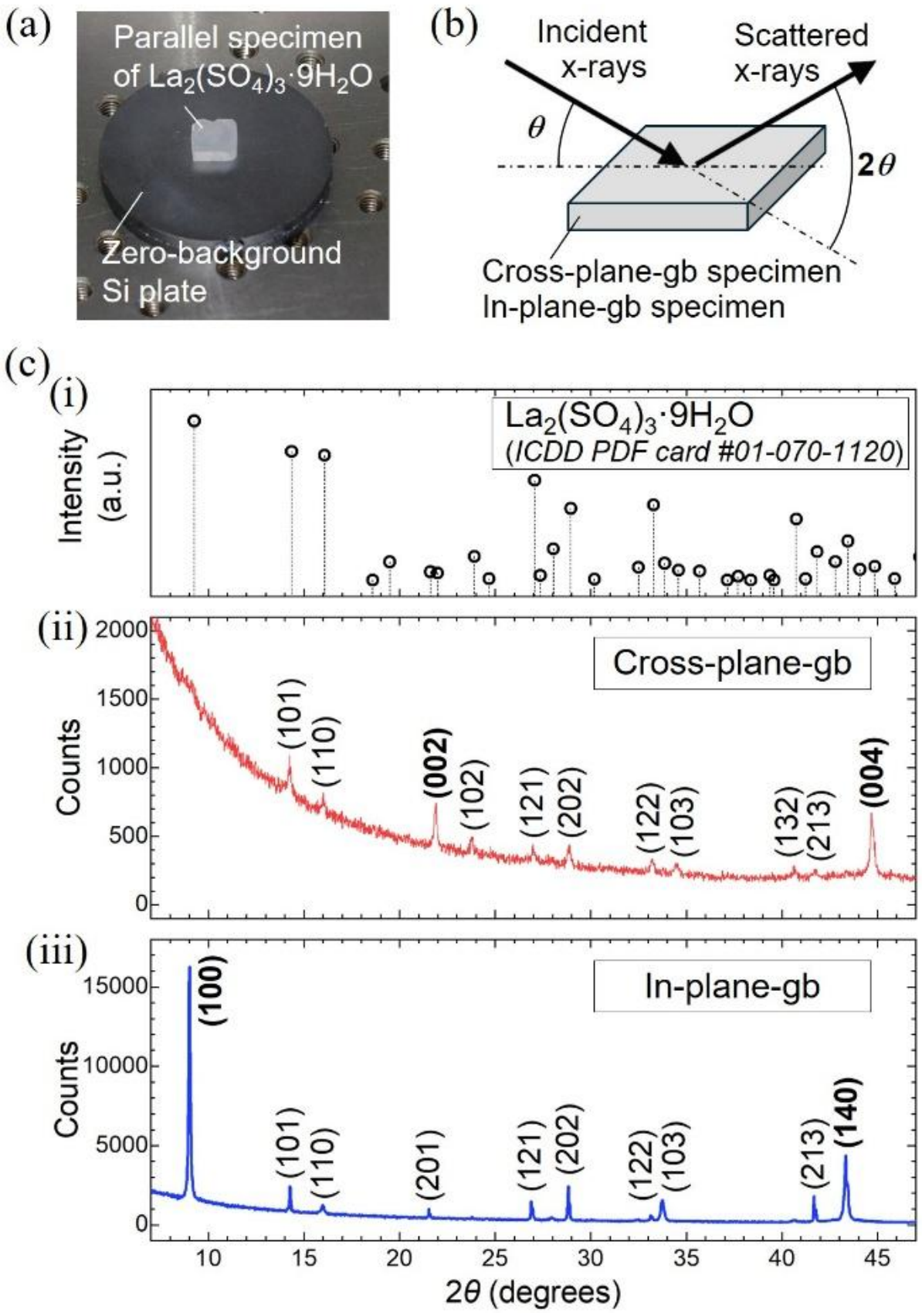

**Fig. 4.** (a) Photograph of the XRD setup with a specimen mounted on a zero-background silicon plate. (b) Schematic of the θ–2θ geometry for plate-shaped specimens. (c) XRD patterns of representative $La_2(SO_4)_3·9H_2O$: (i) ICDD reference [35], (ii) cross-plane-gb specimen, and (iii) in-plane-gb specimen. The diffraction planes of the XRD peaks in panels (ii) and (iii) were indexed to the reference pattern and are indicated in the figure. Diffractions significantly stronger than those in the reference pattern are highlighted in bold, representing preferred orientations of the specimens.

### 3.3 Microstructural observation of cross-plane-gb and in-plane-gb specimens of β-$La_2(SO_4)_3$

To clarify the orientation of grain-boundary alignment in the microstructure of β-$La_2(SO_4)_3$, cross-sectional SEM analyses were conducted on the cross-plane-gb and in-plane-gb specimens after heating at 300 °C to induce dehydration from $La_2(SO_4)_3·9H_2O$ to β-$La_2(SO_4)_3$. Fig. 5 shows the cross-sectional photographs and SEM images of the cross-plane-gb and in-plane-gb specimens of β-$La_2(SO_4)_3$. While the initial $La_2(SO_4)_3·9H_2O$ specimens were translucent, as shown in Fig. 3, the dehydrated specimens appeared white owing to microstructural changes caused by polycrystallization during dehydration [31]. The most prominent feature was the aligned grain boundaries. In the cross-plane-gb specimen, the grain boundaries were predominantly aligned along the cross-plane direction, parallel to the heat-flow direction during LFA (Section 3.4). By contrast, in the in-plane-gb specimen, the grain boundaries lay mostly in-plane and were therefore perpendicular to the heat-flow direction. Fig. 6 compares the XRD patterns of representative cross-plane-gb and in-plane-gb β-$La_2(SO_4)_3$ specimens with that of β-$La_2(SO_4)_3$ powder. The clear differences in relative peak intensities from the powder pattern indicate that both plate specimens exhibit preferred crystallographic orientation. The in-plane-gb specimen shows enhanced h00 reflections, whereas the cross-plane-gb specimen exhibits relatively enhanced hk0 reflections. Because multiple reflections overlap in the 25–26° region, further orientation analysis is difficult based on θ–2θ data alone.

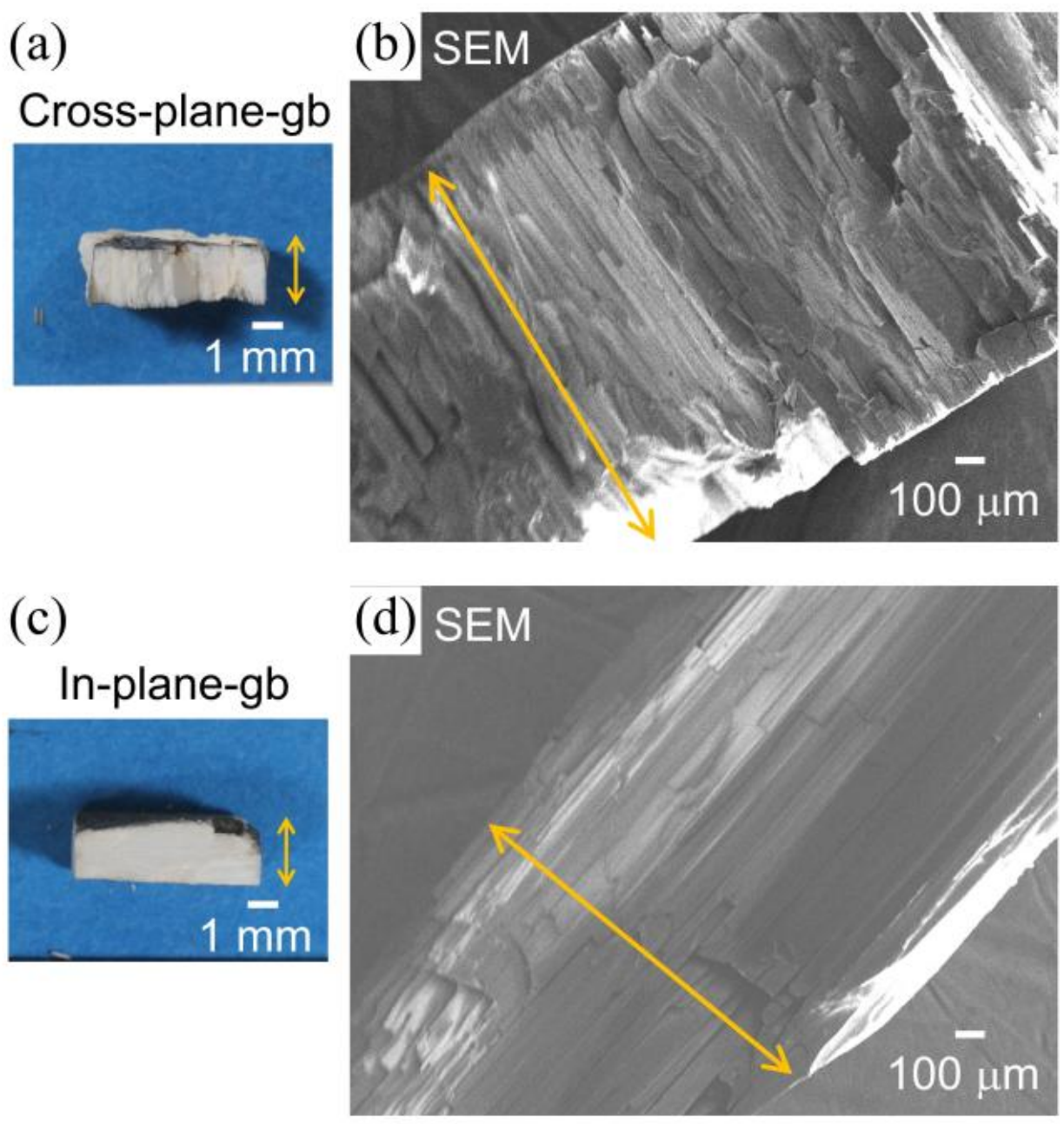

**Fig. 5**. (a) Photograph and (b) SEM image of the cross-plane-gb specimen of β-$La_2(SO_4)_3$. (c) Photograph and (d) SEM image of the in-plane-gb specimen. The SEM images were acquired using EPMA. The orange double-headed arrows indicate the thickness direction for each image, corresponding to the heat-flow direction during LFA measurements.

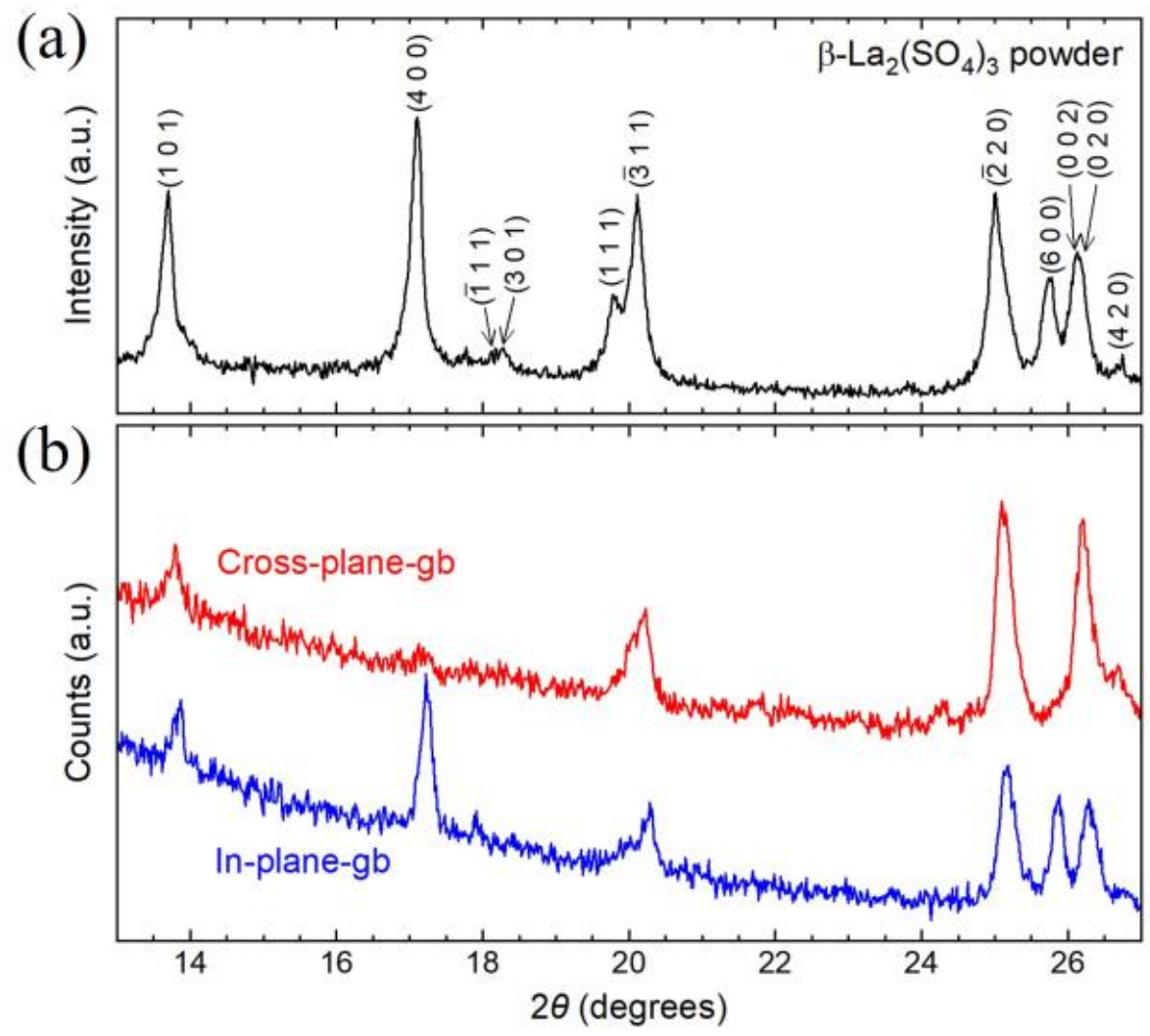

**Fig. 6. (a)** XRD pattern of β-$La_2(SO_4)_3$ powder, prepared by dehydrating $La_2(SO_4)_3{\cdot}9H_2O$ (>98.0%, Wako Pure Chemical) at 300 °C followed by ball milling for 24h. (b) representative cross-plane-gb and in-plane-gb specimens by θ–2θ scan, as shown in Figs. 4(a) and 4(b). The red and blue patterns represent the cross-plane-gb and in-plane-gb specimens, respectively. The diffraction indices are defined assuming a monoclinic crystal system with $\gamma \neq 90°$, based on the monohydrate phase: β-$La_2(SO_4)_3{\cdot}H_2O$ ($x$ = 1) [30].

### 3.4 Thermal diffusivity of cross-plane-gb and in-plane-gb specimens during dehydration

Laser flash analysis (LFA) was conducted on cross-plane-gb and in-plane-gb specimens of $La_2(SO_4)_3{\cdot}9H_2O$. Upon heating from 30 to 300 °C, dehydration proceeded stepwise, first yielding amorphous $La_2(SO_4)_3{\cdot}nH_2O$ and then crystalline β-$La_2(SO_4)_3$ [30]. Fig. 7 shows the thermal diffusivities measured at each stage. Two specimens (#1, #2) were prepared and measured for each specimen type (cross-plane-gb and in-plane-gb) of $La_2(SO_4)_3{\cdot}9H_2O$. For the cross-plane-gb specimen #1, measurements were carried out at 30, 45, 60, 160, 250, and 300 °C. However, the specimen fractured above 160 °C, allowing partial transmission of the Xe lamp, and further measurements were not possible. For specimen #2, measurements were restricted to 30, 60, and 300 °C. In this case, cracks again formed during heating, but they could be repaired using a zirconia-based ceramic adhesive, allowing the measurements to be continued. The influence of this repair was examined. The results are shown in Fig. S2, which indicates that the diffusivity values may have been slightly overestimated, but the effect was less than 8%.

For the initial $La_2(SO_4)_3{\cdot}9H_2O$ phase, the measured diffusivities varied widely, and no statistically significant orientation dependence was identified. By contrast, a clear orientation dependence emerged after dehydration. β-$La_2(SO_4)_3$ derived from cross-plane-gb specimen #2 exhibited higher diffusivity (~0.41 mm²/s), whereas those derived from in-plane-gb specimens showed lower values (0.24–0.30 mm²/s). At 160 °C, thermal diffusivity was significantly reduced owing to the disordered structure, corresponding to the amorphous phase [36]. These data are thermal diffusivities and are not corrected for heat capacity. Therefore, the trends do not necessarily correspond directly to thermal conductivity. It should be noted that only one cross-plane-gb specimen could be successfully measured in this series. Therefore, to ensure reproducibility and verify the orientation dependence, additional measurements of multiple specimens are presented in Section 3.5.

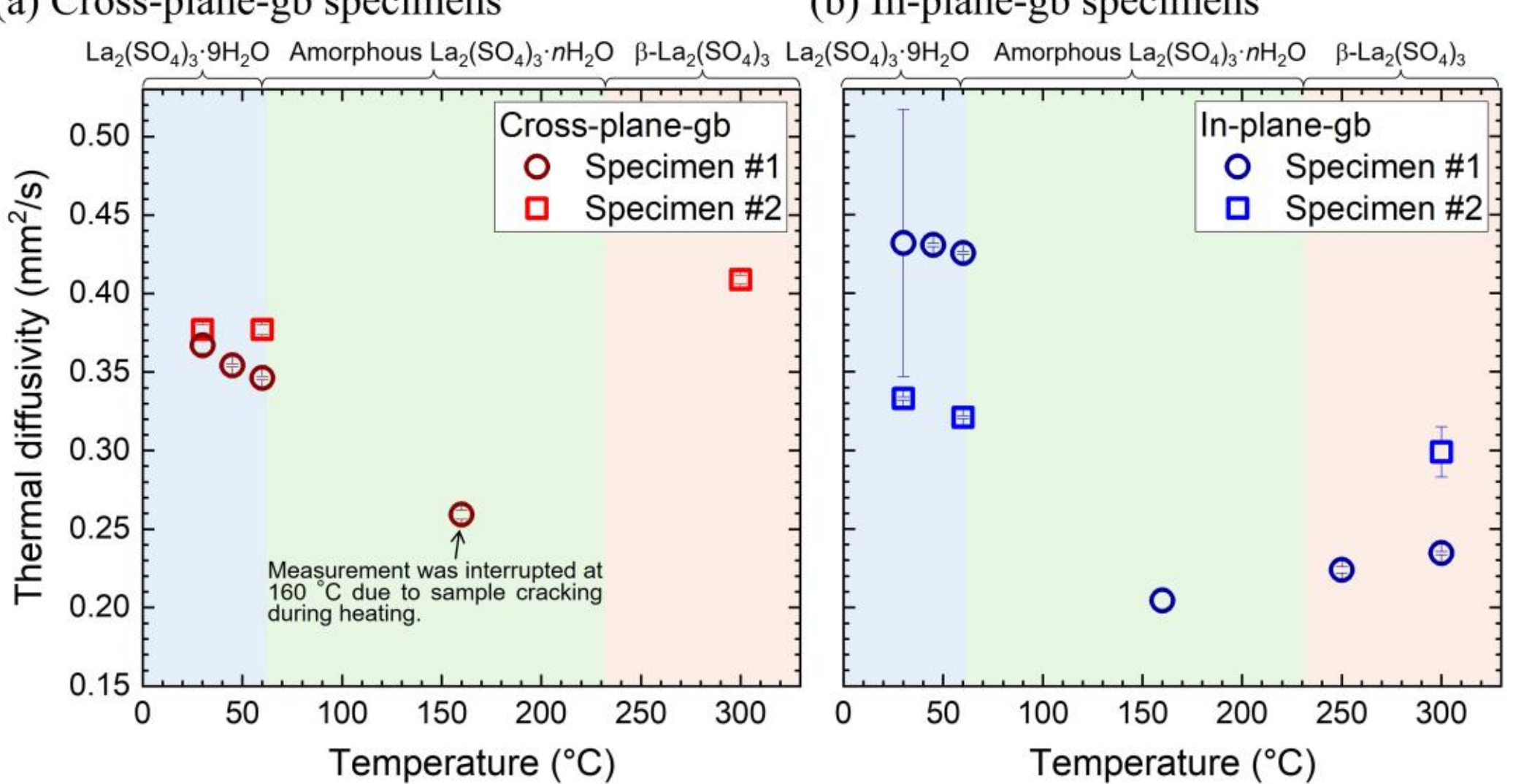

**Fig. 7**. Thermal diffusivity of (a) cross-plane-gb and (b) in-plane-gb specimens of $La_2(SO_4)_3·9H_2O$ measured by LFA under Ar atmosphere. The stability ranges of $La_2(SO_4)_3·9H_2O$, amorphous $La_2(SO_4)_3·nH_2O$, and β-$La_2(SO_4)_3$ are highlighted in blue, green, and red, respectively. Error bars represent the standard deviation obtained from repeated measurements on the same specimen at each temperature.

### 3.5 *Ex situ* verification using pre-dehydrated β-$La_2(SO_4)_3$ specimens

To confirm the orientation effect shown in Fig. 7, we performed additional LFA measurements on pre-dehydrated β-$La_2(SO_4)_3$ specimens prepared *ex situ* from cross-plane-gb and in-plane-gb specimens of $La_2(SO_4)_3·9H_2O$ precursors. Because the specimens were prone to fracture during dehydration and subsequent handling, their side surfaces were reinforced by inorganic putty or silicone rubber prior to furnace treatment. The side surfaces of the plate-shaped specimen are located outside the heating area of the LFA; therefore, the reinforcement materials are considered to have negligible influence on the measured thermal diffusivity. Multiple cross-plane-gb and in-plane-gb specimens of β-$La_2(SO_4)_3$ and the powder compact of β-$La_2(SO_4)_3$ were prepared and measured. Fig. 8 summarizes the LFA measurement results. A consistent orientation dependence was observed.

Regarding measurement reliability, xenon flash irradiation may in principle induce dehydration in hydrated or partially hydrated states, which could affect the measured thermal diffusivity [37]. However, in the present study, the temperature rise during the LFA measurement is expected to be on the order of a few Kelvin, as is typical for LFA experiments. This is supported by the fact that the thermal diffusivity of a 2-mm-thick Pyroceram reference measured under identical conditions agrees well with the literature value. Therefore, the measurements can be regarded as having been conducted under standard LFA conditions with a sufficiently small temperature rise. Since dehydration of $La_2(SO_4)_3$ occurs above approximately 80 °C [30], its influence is expected to be negligible. Furthermore, although local heating near the irradiated surface may exceed this temperature, such effects are limited to a shallow region and have a negligible impact on the evaluated thermal diffusivity, which is governed by bulk heat transport in the 1–2 mm thick sample.

Specifically, the cross-plane-gb specimens of β-$La_2(SO_4)_3$ showed higher thermal diffusivities than the in-plane-gb specimens, whose values were comparable to that of the randomly oriented β-$La_2(SO_4)_3$ powder compact. This result clearly indicates that heat transport in β-$La_2(SO_4)_3$ is strongly influenced by microstructural orientation and is consistent with the trend observed in the in situ measurements (Section 3.4). However, the absolute values of thermal diffusivity differ from those in Fig. 7. In the ex situ tests (Fig. 8), the $La_2(SO_4)_3{\cdot}9H_2O$ precursors were dehydrated in a furnace and subsequently exposed to ambient conditions, under which rehydration to β-$La_2(SO_4)_3{\cdot}H_2O$ ($x$ = 1) is expected. In contrast, the values in Fig. 7 correspond to fully dehydrated β-$La_2(SO_4)_3$ ($x$ = 0). The difference in heat capacity associated with the hydration state can partially account for the reduced thermal diffusivity in the hydrated phase; however, the observed decrease appears too large to be explained by this effect alone. One possible explanation is a change in interfacial thermal conductance between β-$La_2(SO_4)_3{\cdot}xH_2O$ and β-$La_2(SO_4)_3$. Similar hydration-dependent variations in interfacial thermal resistance have been reported in metal–organic framework (MOF) systems [38].

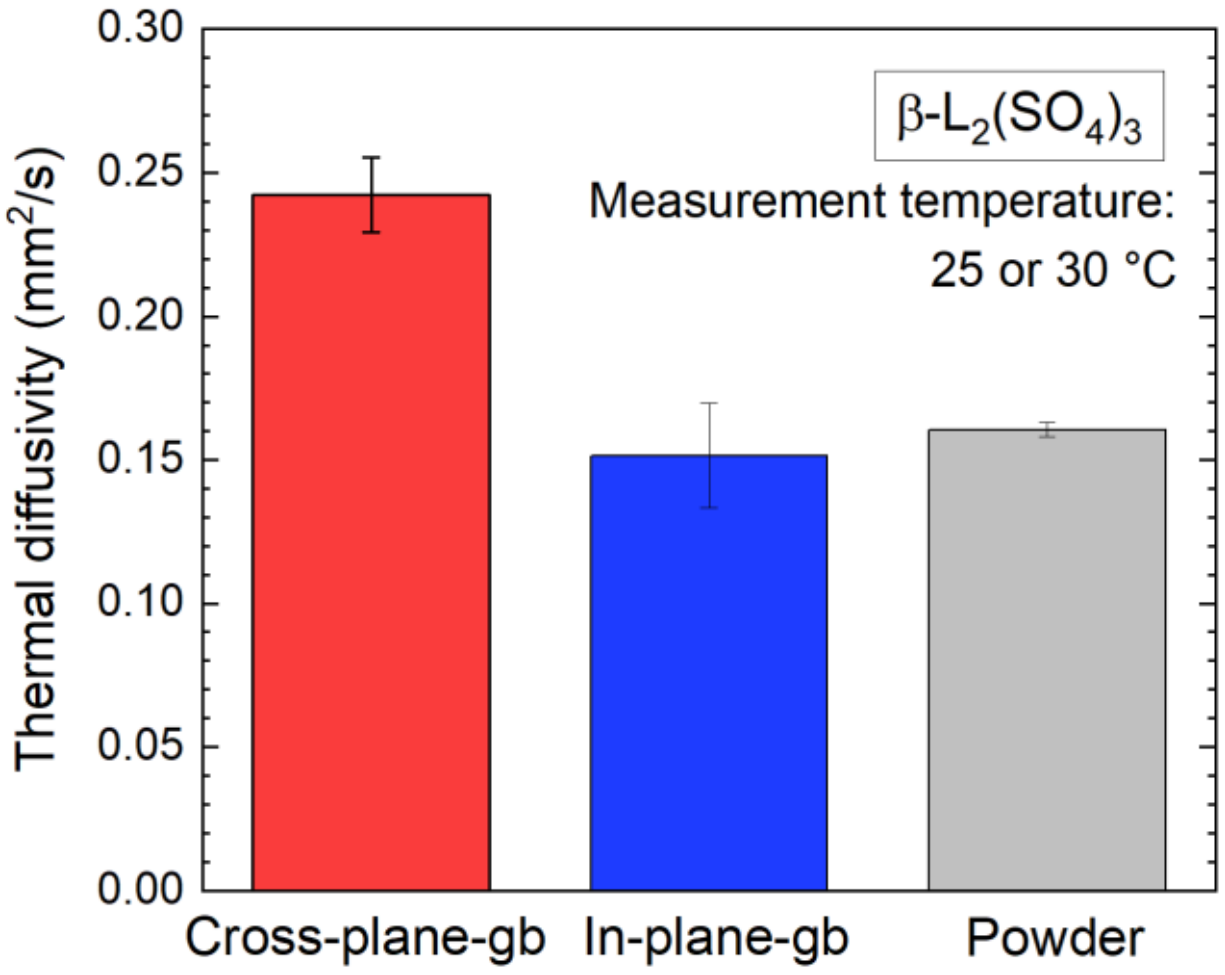

**Fig. 8.** Thermal diffusivity of pre-dehydrated β-$La_2(SO_4)_3$ measured by LFA at 25 or 30 °C for cross-plane-gb and in-plane-gb specimens, together with a β-$La_2(SO_4)_3$ powder compact. All β-phase specimens were obtained by *ex situ* dehydration in an electric furnace under air from the corresponding $La_2(SO_4)_3{\cdot}9H_2O$ precursors. Bars show the mean value across $n$ = 3 (cross-plane-gb), $n$ = 4 (in-plane-gb), and $n$ = 2 (powder compact) distinct specimens. Error bars indicate the standard deviation. The dehydration conditions, relative densities, and thermal diffusivities of each specimen are provided in Table S1.

### 3.6 Implications for heat transport design in TCHS materials

The *ex situ* specimens exhibited the same preferred crystallographic orientation trend as that shown in Fig. 6, as evidenced by a systematic difference in the XRD peak-area ratio $A_{(400)}/A_{(101)}$ (Fig. S3) between the cross-plane-gb and in-plane-gb specimens. These results suggest that the two specimen types differ in both microstructural and crystallographic orientations. Accordingly, controlling the specimen orientation is an effective route to enhance heat transport in β-$La_2(SO_4)_3$. However, it remains unclear whether the improved heat transport in the cross-plane-gb specimens originates solely from preferential microstructural alignment or is also influenced by intrinsic thermal-conductivity anisotropy associated with crystallographic orientation. Since grain boundaries are expected to act as thermal resistance interfaces, the reduced number of grain-boundary crossings along the heat-flow direction in the cross-plane-gb configuration, where heat flow is parallel to the grain-boundary alignment, is likely to play a major role in enhancing heat transport, as schematically illustrated in Fig. 9. Nevertheless, further investigation is required to decouple the respective contributions of microstructural and crystallographic effects.

This strategy of microstructural orientation control is not limited to β-$La_2(SO_4)_3$. Similar oriented microstructures are observed in several TCHS materials, such as $CaSO_4$ (Fig. S4), $MgSO_4$ (Fig. S5), and $Y_2(SO_4)_3$ [39] , which develop anisotropic features during crystal growth or dehydration processes. These materials may also exhibit improved heat transport through orientation control. Although not universally applicable, this approach may provide a useful strategy for enhancing heat transport in TCHS materials.

In addition, from a practical perspective, the present strategy can be extended to macroscopic material design. For example, polycrystalline bodies with controlled microstructural alignment may first be fabricated, and then assembled into larger structures by bonding them using thermal interface materials (TIMs). Such oriented bodies could also be directly integrated onto heat-transfer substrates, for instance in fin-type thermochemical storage systems [40], to facilitate efficient heat extraction from the storage material. Experimental validation of such system-level implementations remains a subject for future work.

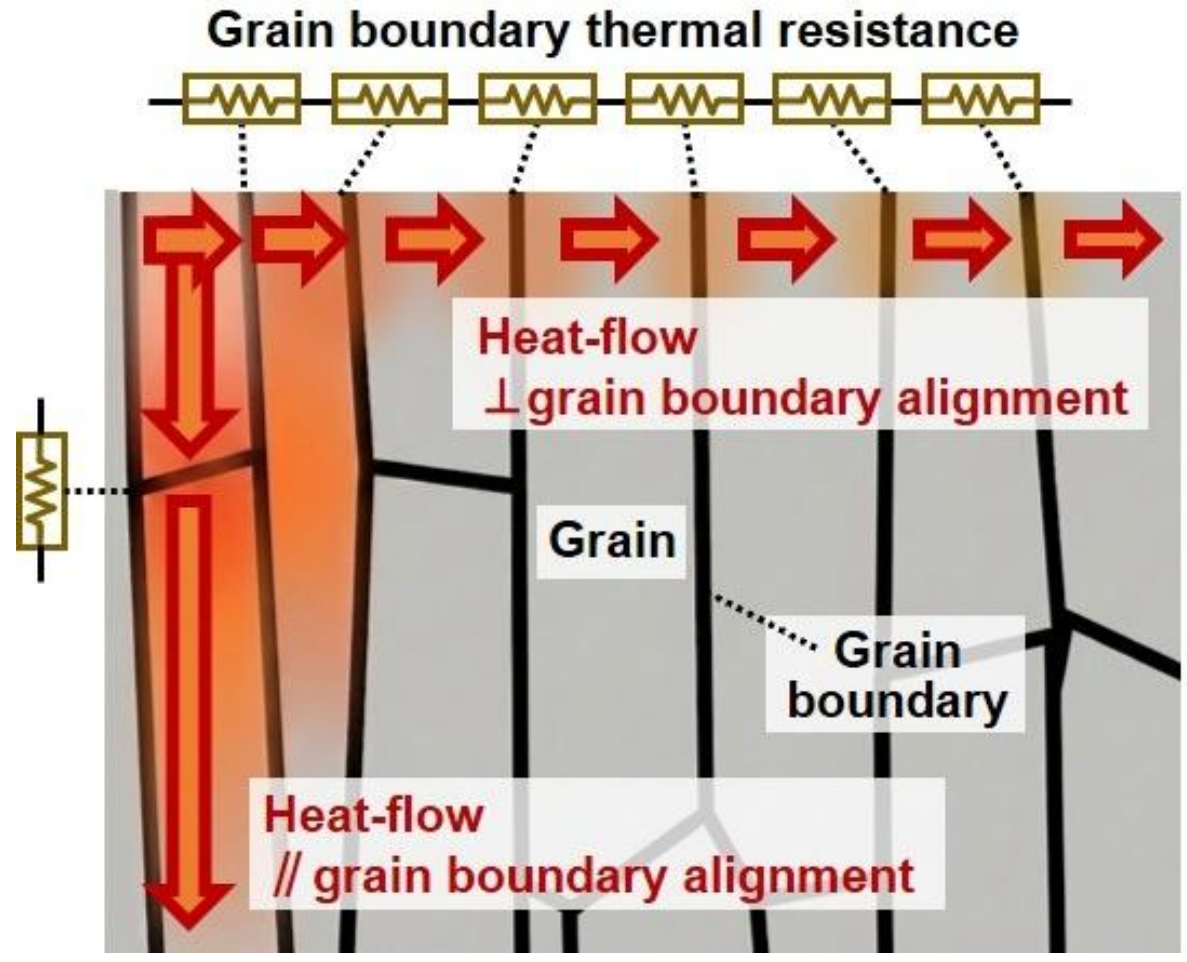

**Fig. 9.** Schematic illustration of heat transport in β-$La_2(SO_4)_3$ for different grain-boundary orientations. When heat flow is perpendicular to the grain-boundary alignment, frequent grain-boundary crossings act as thermal resistances connected in series, resulting in higher total thermal resistance. In contrast, when heat flow is parallel to the grain-boundary alignment, fewer grain-boundary crossings lead to lower total thermal resistance. These configurations correspond to the in-plane-gb and cross-plane-gb measurements in laser flash analysis (LFA), respectively.

## Conclusion

This study demonstrates that the heat-transport efficiency of β-$La_2(SO_4)_3$ strongly depends on its microstructure, and that establishing efficient transport pathways through microstructural orientation control is an effective strategy for enhancing heat transport. Plate-shaped specimens were prepared from highly oriented rod-like $La_2(SO_4)_3 \cdot 9H_2O$ crystals with centimeter-scale lengths. The crystals were sectioned such that the through-thickness direction of the plates was either parallel or perpendicular to this longitudinal direction, yielding two specimen types—cross-plane-gb and in-plane-gb—based on the expected grain-boundary orientation in β-$La_2(SO_4)_3$ after dehydration.

Microstructural observations of the dehydrated β-$La_2(SO_4)_3$ confirmed that the grain boundaries were predominantly aligned along the through-thickness (cross-plane) direction in the cross-plane-gb specimens, whereas they lay mainly within the plate plane (in-plane) in the in-plane-gb specimens. LFA revealed that this orientation difference led to a clear anisotropy in thermal diffusivity. During *in situ* dehydration, the thermal diffusivity of β-$La_2(SO_4)_3$ derived from the cross-plane-gb specimens reached 0.41 $mm^2/s$, whereas that derived from the in-plane-gb specimens showed lower values of 0.24–0.30 $mm^2/s$. This orientation dependence was consistently reproduced in *ex situ* tests, where the cross-plane-gb specimens showed a higher thermal diffusivity (0.24 $mm^2/s$) than the in-plane-gb specimens (0.14 $mm^2/s$). The latter was comparable to that of a randomly oriented β-$La_2(SO_4)_3$ powder compact. These results establish the orientation control as a promising, additive-free route to improve heat transport in several TCHS materials. Although this study focused on β-$La_2(SO_4)_3$, similar oriented microstructures are expected to form in other systems during the crystal growth or dehydration processes. In addition, this approach may be combined with conventional strategies, such as the incorporation of thermally conductive additives, to further improve heat-transfer performance. The development of such design concepts, together with their implementation in practical material forms, represents an important direction for future work.

**Acknowledgements**

The authors thank Prof. Tetsuya Uda (Kyoto University), the principal investigator of this project, for providing access to experimental facilities and for his valuable support and guidance. We also thank Prof. Yasushi Hamanaka (Nagoya Institute of Technology) and Assoc. Prof. Hidetoshi Miyazaki (Nagoya Institute of Technology) for access to experimental facilities. We would like to thank Editage (www.editage.jp) for English language editing.

**Funding**

This work was supported by a JSPS Grant-in-Aid for Young Scientists (B) (Grant Number 17K17821), Grant-in-Aid for JSPS Fellows (Grant Number 19J15085), Grant-in-Aid for Young Scientists（Grant Number 23K13818）.

**Declaration of generative AI and AI-assisted technologies in the manuscript preparation process.**

During the preparation of this work the authors used ChatGPT and Perplexity in order to prepare the manuscript. After using this tool, the authors reviewed and edited the content as needed and take full responsibility for the content of the published article.

**Author contributions: CRediT**

**Kunihiko Shizume:** Funding acquisition, Investigation, Methodology, Writing – original draft

**Naoyuki Hatada:** Conceptualization, Funding acquisition, Investigation, Writing - Review & Editing, Supervision

# Supplementary Material

## Microstructural Control and Heat Transport Enhancement in Lanthanum Sulfate for Thermochemical Heat Storage

**Kunihiko Shizume[1,2], Naoyuki Hatada[1]**

**[1] Kyoto University**

**[2] Nagoya Institute of Technology**

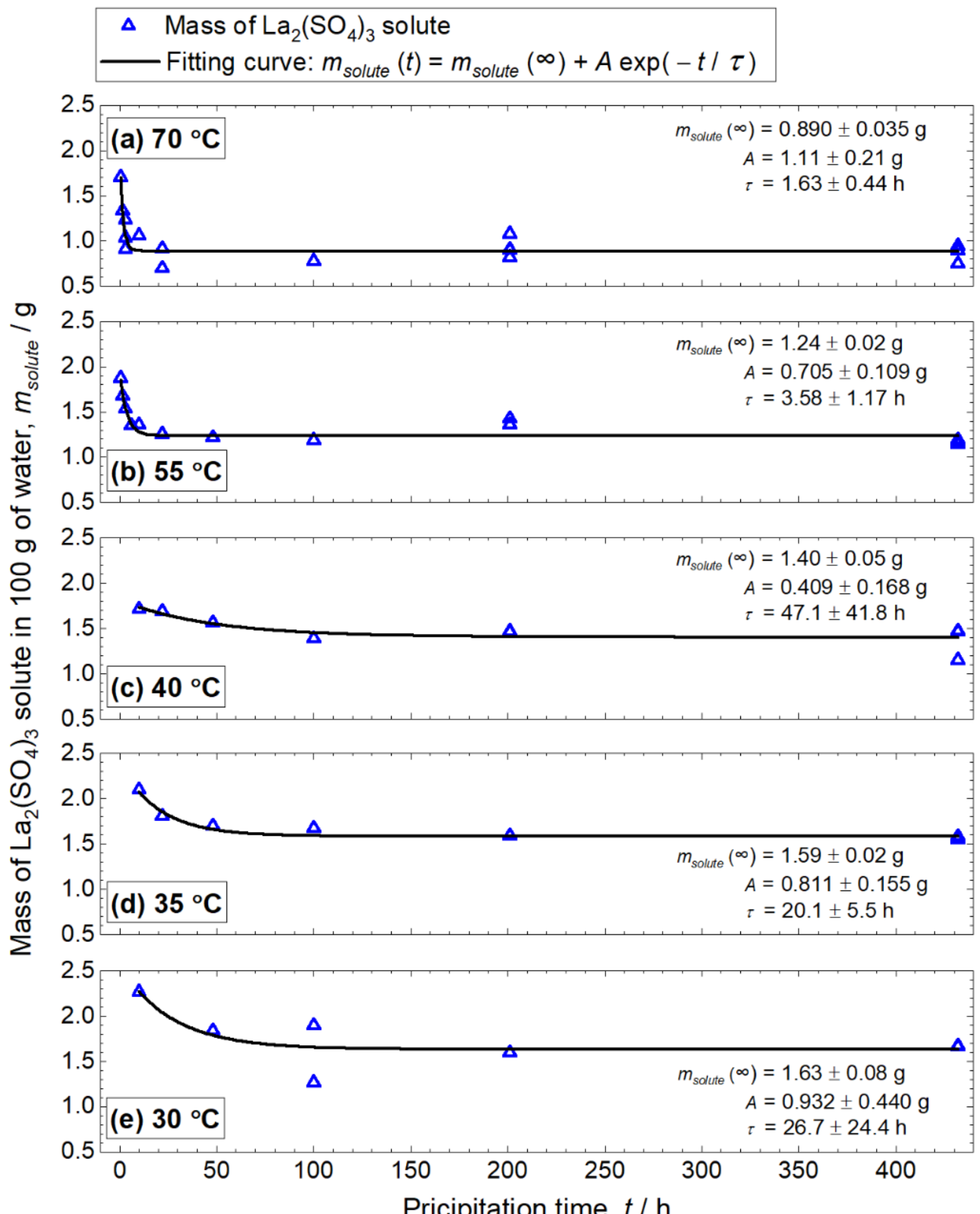

**Fig. S1.** Time evolution of the residual mass of $La_2(SO_4)_3$, determined by thermogravimetric analysis (TGA) at varying holding temperatures (30, 35, 40, 55, and 70 °C). Each panel displays the temporal change in solute mass at the specified temperature, with data collected after 10, 48, 100, 201, and 432 h of precipitation. The TGA measurements were performed under flowing dry Ar, with the samples heated and maintained at 80 °C. The experimental data were fitted to extract the equilibrium solute content. The time-dependent change in the residual mass of $La_2(SO_4)_3$, $m_{solute}(t)$, was fitted using an exponential decay function of the form $m_{solute}(t) = m_{solute}(\infty) + A\exp(-t/\tau)$. Here, $m_{solute}(\infty)$ represents the equilibrium solute mass at saturation, corresponding to the solubility; $A$ is the amplitude associated with the initial supersaturation; $t$ is the elapsed time; and $\tau$ is the characteristic time constant for the precipitation process. This function describes the approach of the solute mass toward equilibrium, assuming first-order kinetics for the decrease in supersaturation. The time-dependent decrease in solute mass was fitted with a monoexponential decay function to determine the equilibrium solute amount, corresponding to the solubility of $La_2(SO_4)_3$ in water.

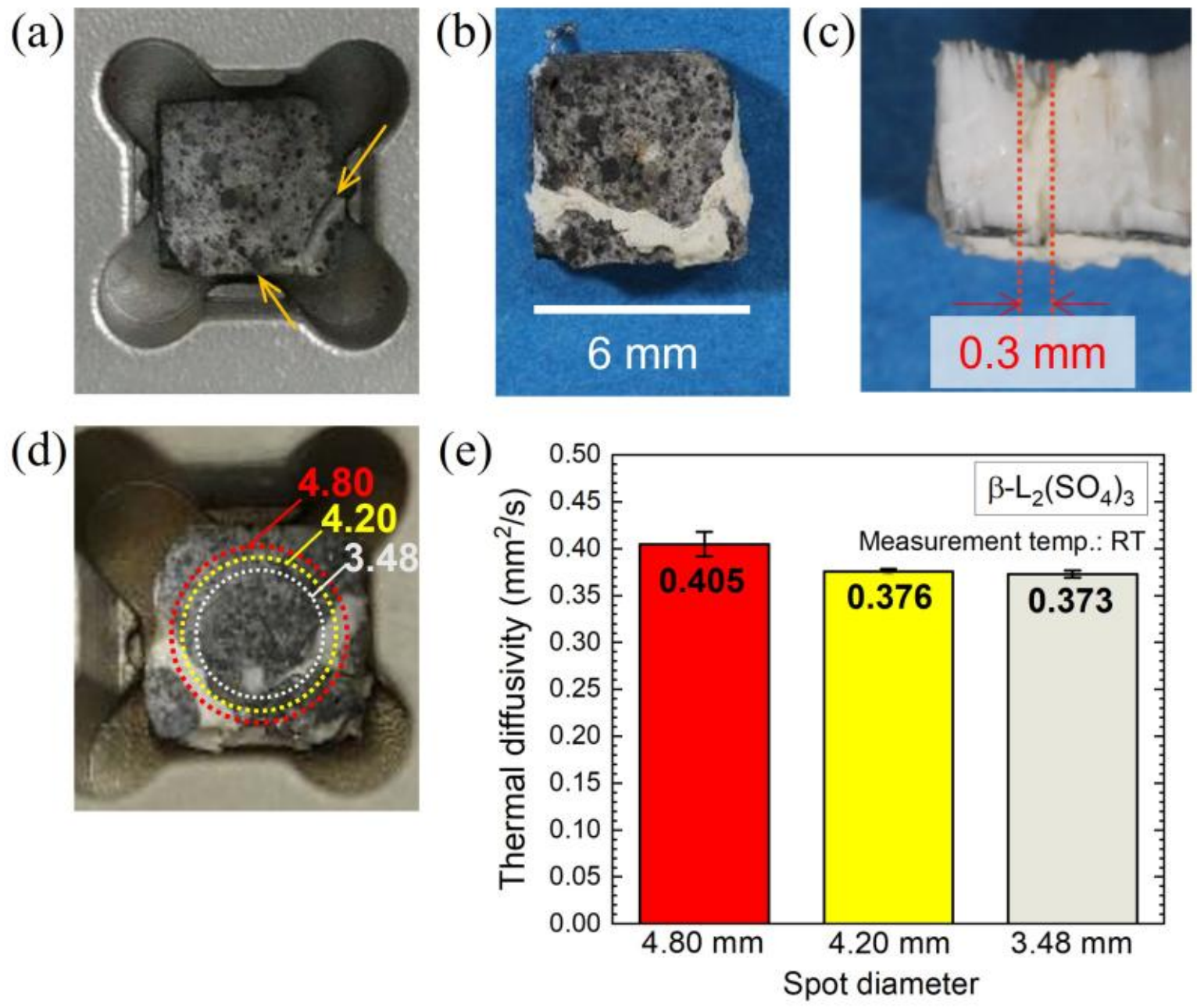

**Fig. S2**. Cracking and repair of cross-plane-gb specimen #2 during LFA measurement, and the effect of the repair material on thermal diffusivity. (a) Cross-plane-gb specimen #2 after heating above 160 °C in the LFA test of Fig. 7. Cracks appeared at the positions marked by orange arrows. (b) Specimen after repair with zirconia-based ceramic adhesive. The white-colored regions correspond to the adhesive. (c) Cross-section of the repaired specimen. The adhesive layer is thin, and the thickness was estimated to be ~0.3 mm. (d) After completing the LFA measurements and heating up to 300 °C (Fig. 7), additional room-temperature LFA measurements were carried out using different spot diameters of the Xe lamp. The red, yellow, and gray circles indicate the estimated heating areas for spot diameters of 4.80, 4.20, and 3.48 mm, respectively. With a 4.80 mm spot size, the heating area overlapped with the adhesive region, whereas for 4.20 and 3.48 mm, no significant overlap was expected. Note that the measurements in Fig. 7 were performed with a 4.80 mm spot. (e) Thermal diffusivity measured for each spot size. Error bars show the standard deviation (SD) of three repeated measurements. The thermal diffusivity values for 4.20 and 3.48 mm were comparable, while 4.80 mm gave slightly higher thermal diffusivities, consistent with the adhesive overlap shown in (d). This indicates that the adhesive can cause an apparent increase in diffusivity; however, the overestimation was limited to ~8%. Therefore, the difference in diffusivity between cross-plane-gb and in-plane-gb specimens of β-$La_2(SO_4)_3$ at 300 °C in Fig. 7 cannot be attributed solely to the presence of cement and represents a significant intrinsic orientation effect.

**Table S1**. Dehydration conditions, relative densities, and thermal diffusivities of each specimen shown in Fig. 8.

| Type | # | Preheating treatment | | Reinforcement material[a] | LFA meas. temp. (°C) | Thermal diffusivity ($mm^2/s$) | Relative density[b] (%) |
|---|---|---|---|---|---|---|---|
| | | Temperature (°C) | Time (min) | | | | |
| Cross-plane-gb β-$La_2(SO_4)_3$ | 1 | 300 | 30 | Inorganic putty | 25 | 0.259±0.001 | 73 |
| | 2 | 270 | 90 | Silicone rubber | 25 | 0.227±0.001 | 68 |
| | 3 | 270 | 90 | Silicone rubber | 25 | 0.241±0.001 | 68 |
| In-plane-gb β-$La_2(SO_4)_3$ | 1 | 300 | 30 | Inorganic putty | 25 | 0.140±0.003 | 73 |
| | 2 | 300 | 90 | Silicone rubber | 25 | 0.177±0.007 | 61 |
| | 3 | 270 | 90 | Silicone rubber | 25 | 0.129±0.001 | 68 |
| | 4 | 270 | 90 | Silicone rubber | 25 | 0.159±0.000 | 73 |
| Powder compact β-$La_2(SO_4)_3$ | 1 | 300 | 90 | - | 30 | 0.163±0.001 | 60 |
| | 2 | 300 | 90 | - | 25 | 0.158±0.003 | 60 |

[a] Silicone rubber: RTV silicone rubber (ShinEtsu Silicone, KE-3418); Inorganic putty: Refractory putty (Cemedine, HJ-112).

[b] The relative densities of the specimens were determined from their dimensions and mass.

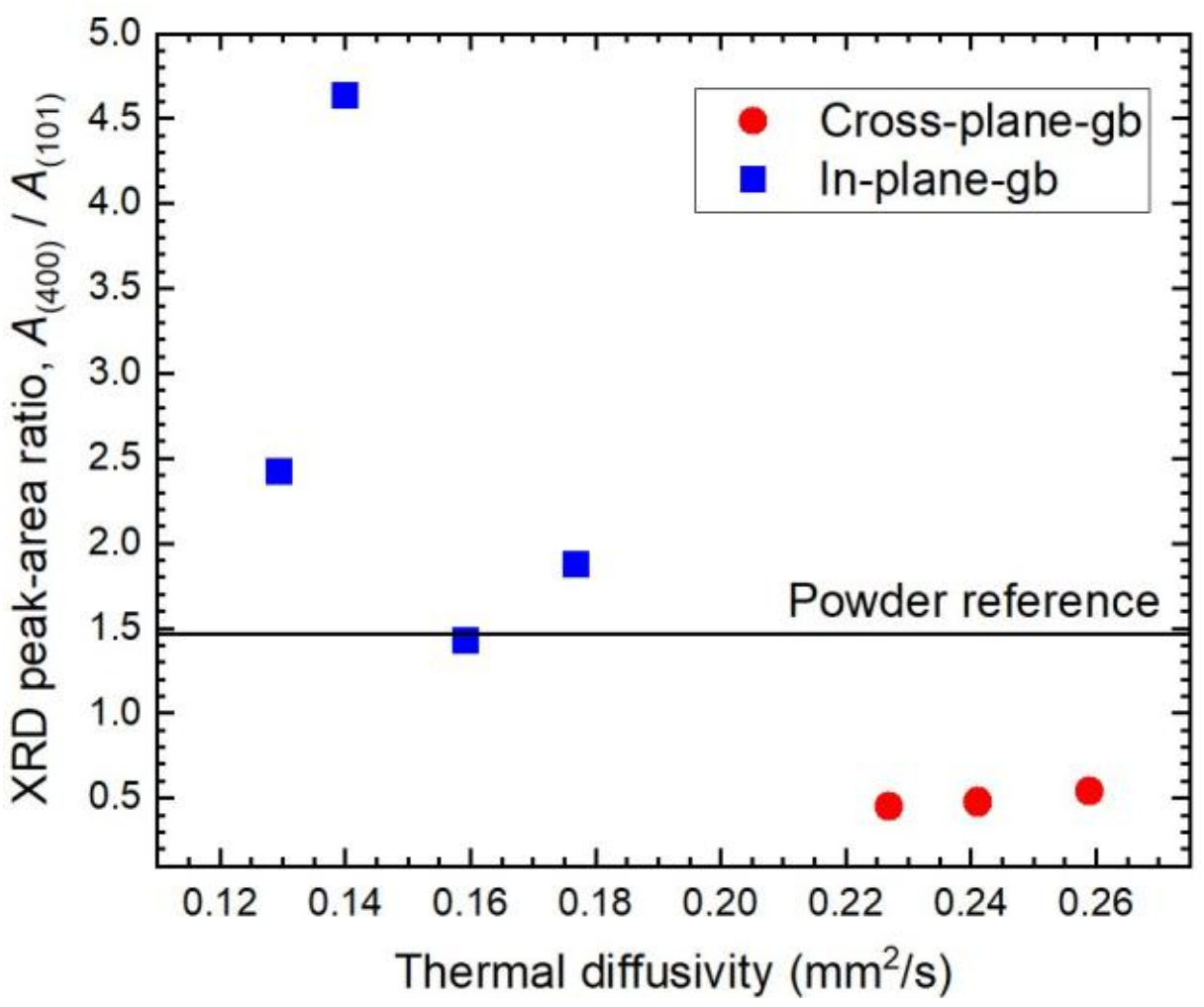

**Fig. S3.** Relationship between thermal diffusivity and the XRD peak-area ratio $A_{(400)}/A_{(101)}$ for cross-plane-gb and in-plane-gb β-$La_2(SO_4)_3$ specimens shown in Fig. 8. $A_{(hkl)}$ denotes the baseline-corrected peak area obtained by summing the signal within ±0.35° of the peak-top position. The horizontal line indicates the reference value obtained for the β-$La_2(SO_4)_3$ powder shown in Fig. 6(a) and Table S1. The (101) peak area was comparable across all specimens, regardless of orientation. Therefore, it was used as an internal reference to normalize the (400) peak area.

The in-plane-gb specimens exhibited $A_{(400)}/A_{(101)}$ beyond powder reference, whereas the cross-plane-gb specimens show below, suggesting a different preferred orientation, consistent with the trend in the θ–2θ patterns in Fig. 6. A tendency is also observed that stronger h00 preference corresponds to lower thermal diffusivity. As discussed in the main text, however, the present data do not allow us to separate the contributions of preferential microstructural alignment and intrinsic thermal-conductivity anisotropy.

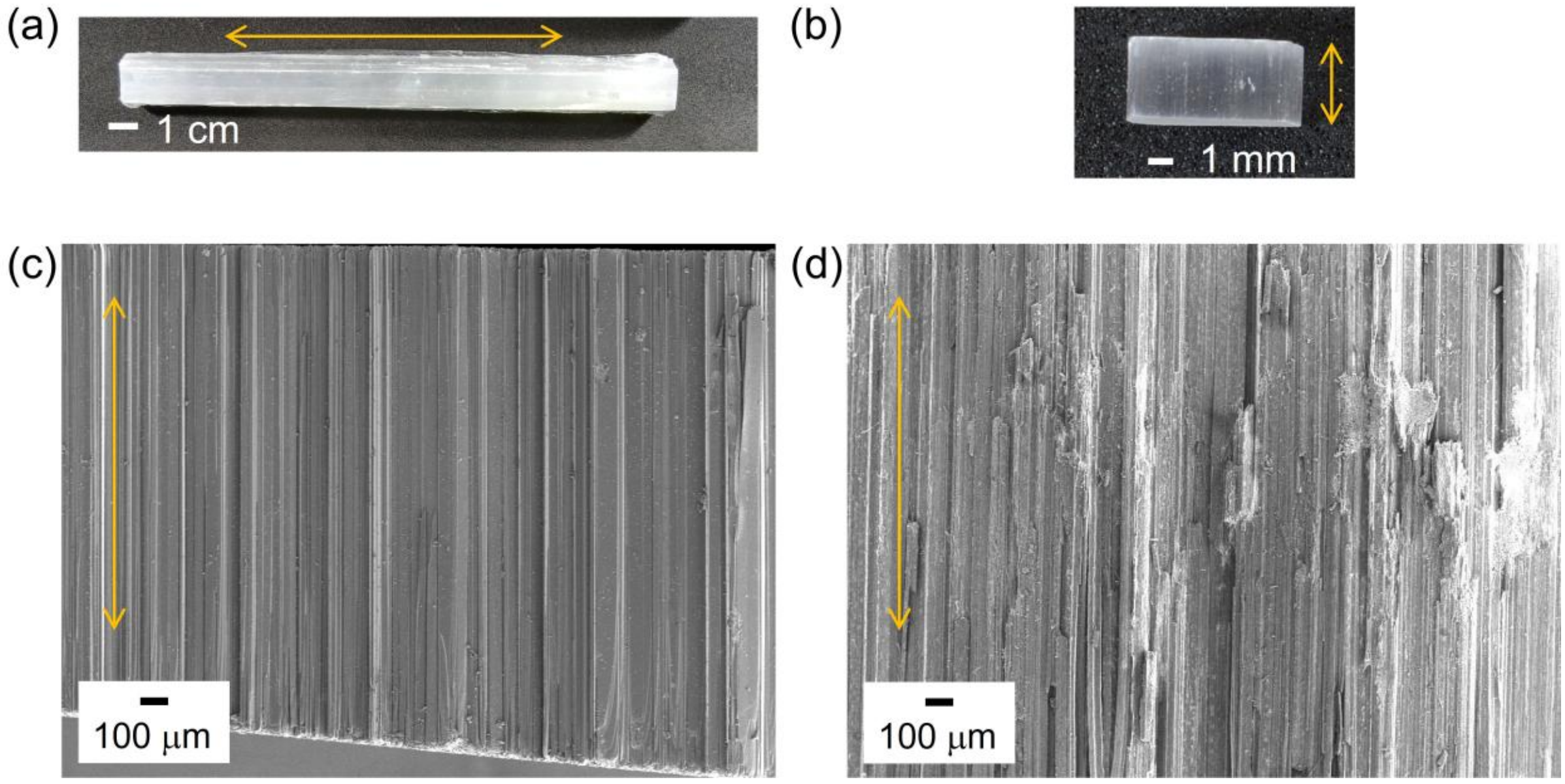

**Fig. S4.** Microstructure of $CaSO_4 \cdot nH_2O$. The dehydration/hydration reaction between $CaSO_4 \cdot 0.5H_2O$ and $CaSO_4$ is considered promising for thermochemical heat storage [S1].

(a) A large polycrystalline body of $CaSO_4 \cdot 2H_2O$ (selenite), a natural mineral from Morocco, commercially purchased.

(b) Cross-sectional image of a piece of the oriented polycrystalline $CaSO_4 \cdot 2H_2O$.

(c) SEM image of the cross section of the oriented $CaSO_4 \cdot 2H_2O$ polycrystal, obtained using a JSM-IT210 (JEOL).

(d) SEM image of the cross section of the oriented $CaSO_4 \cdot 0.5H_2O$ polycrystal after dehydration at 140 °C, obtained using a JSM-IT200 (JEOL).

The orange arrows indicate a common orientation direction in each panel. As shown in (c), the $CaSO_4 \cdot 2H_2O$ precursor already exhibits a oriented microstructure, which is likely formed during natural crystal growth. This orientation is preserved after dehydration to $CaSO_4 \cdot 0.5H_2O$ at 140 °C. These results suggest that preparing high-quality oriented $CaSO_4 \cdot 2H_2O$ precursors could enable the fabrication of polycrystalline materials with enhanced heat transport.

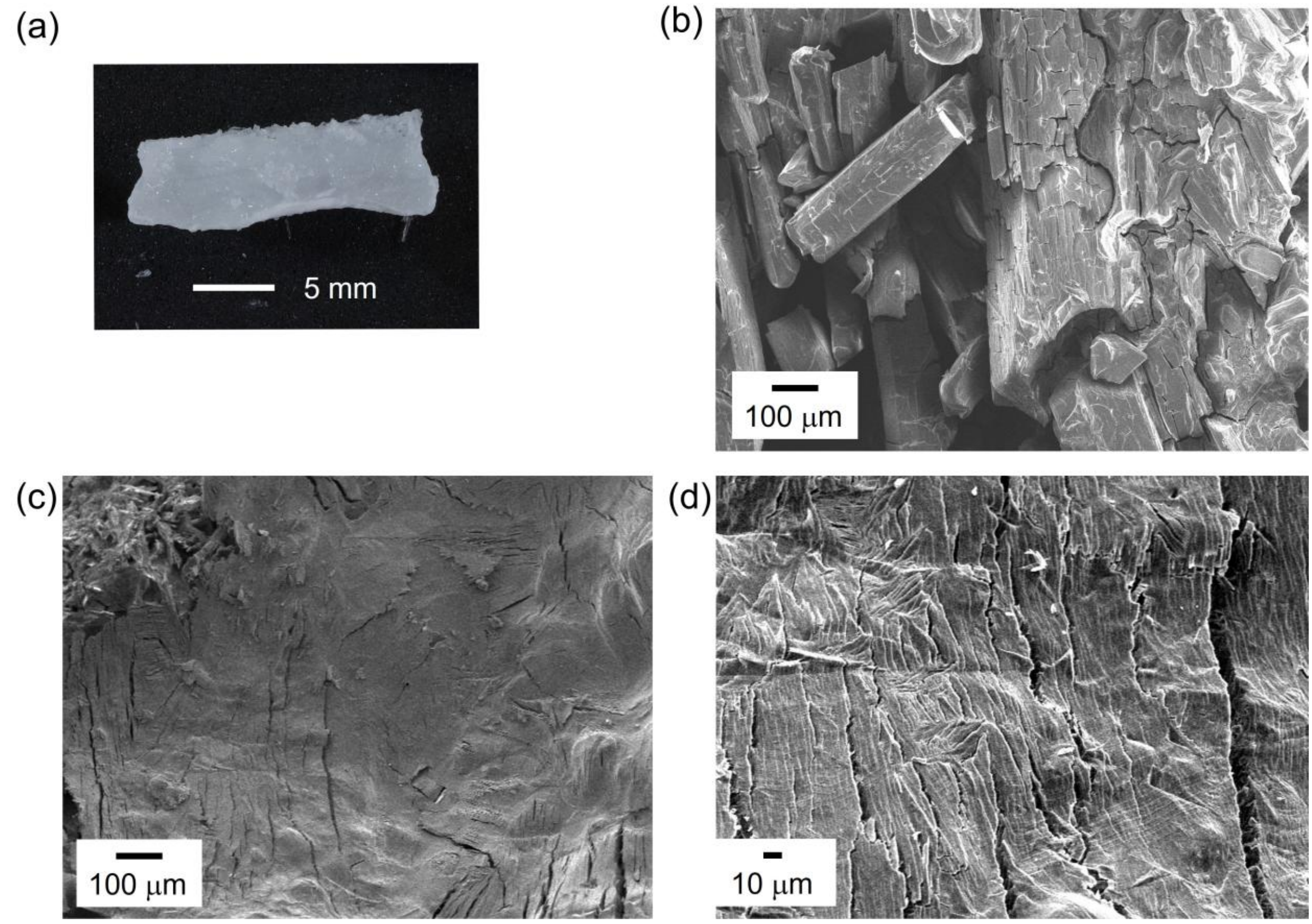

**Fig. S5.** Microstructure of $MgSO_4 \cdot nH_2O$, considered promising for thermochemical heat storage [S2].

(a) A representative photograph of a large polycrystalline body of $MgSO_4 \cdot 7H_2O$ synthesized and grown in the saturated aqueous solution.

(b) SEM image of the cross section of $MgSO_4 \cdot 7H_2O$.

(c) SEM image of the cross section of $MgSO_4$ after dehydration at 300 °C, obtained using a JSM-IT200 (JEOL).

(d) Magnified view of the same sample as in (c).

As shown in (a) and (b), the obtained polycrystalline body exhibits low density and a lower degree of orientation compared to β-$La_2(SO_4)_3$ and $CaSO_4$. In addition, dehydration from $MgSO_4 \cdot 7H_2O$ is accompanied by significant changes in the external dimensions. Therefore, the relationship between crystal orientation and the resulting microstructure is difficult to establish. Nevertheless, locally oriented regions are observed after dehydration, as shown in (c) and (d). These results suggest that improved heat transport could potentially be achieved through orientation control by optimizing the precursor preparation and dehydration conditions.